# Eric Strach: four decades of detailed synoptic solar observations (1969-2008)

V.M.S. Carrasco[1,2], J.M. Vaquero[2,3], V.M. Olmo-Mateos[3]

[1] Departamento de Física, Universidad de Extremadura, 06071 Badajoz, Spain [e-mail: vmscarrasco@unex.es]

[2] Instituto Universitario de Investigación del Agua, Cambio Climático y Sostenibilidad (IACYS), Universidad de Extremadura, 06006 Badajoz, Spain

[3] Departamento de Física, Universidad de Extremadura, 06800 Mérida, Spain

**Abstract:** Eric H. Strach (1914-2011) studied medicine at University of Prague and graduated in 1938. Strach dedicated a great part of his life to astronomy becoming a consistent and meticulous observer. He joined the Liverpool Astronomical Association and British Astronomical Association during the 1960s and obtained two recognitions as proof of his great work in solar physics: the BAA's Merlin Medal and Gift in 1999 and Walter Goodacre Medal and Gift, ten years later. Strach recorded four decades (1969-2008) of systematic solar records in his observation notebooks although he started his observations from the late 1950s. In this work, we document the valuable effort made by Strach in getting four decades of solar records and the importance of this kind of long observation series for studies of space weather and climate. We present the sunspot group number series according to Strach's data and a long observation series of prominences recorded by Strach.

**Key Points**: (1) Eric H. Strach recorded solar observations (sunspots, prominences, plages, filaments, and solar flares) during the period 1969–2008; (2) Strach joined the Liverpool Astronomical Association and British Astronomical Association and was a meticulous observer; (3) Strach's data can be consulted at HASO website (http://haso.unex.es/)

## 1. Introduction

Sunspots on the solar disc or variations in the galactic cosmic-ray intensity measured from cosmogenic radionuclides stores on the Earth are two examples of multiple manifestations of solar activity (Usoskin *et al*., 2002; Muscheler *et al*., 2016; Usoskin, 2017; Barnard *et al*., 2018). Space weather and climate have both general and scientist due to the impact of solar activity on the entire heliosphere and, in particular, on our space environment. For example, space weather effects can cause disturbances in communication systems and induced currents in power lines producing problems in the electricity supply (Ridley *et al*., 2018; Saiz *et al*., 2018). The morphology of sunspot active regions let us know the property of flare-producing active regions (Toriumi *et al*., 2017; Toriumi and Takasao, 2017). The sunspot drawings and photographs have made significant contributions for the understanding of such active regions associated with extreme space weather events (Ishii *et al*., 1998; Cliver and Keer, 2012; Lefèvre *et al*., 2016; Hayakawa *et al*., 2017, 2018, 2019; Love, 2018). Thus, it is fundamental long-term sunspot observations are available for the community of space weather studies in order to continue improving our understanding and prediction about solar activity.

The current knowledge about space climate and weather has been achieved due to systematic sunspot observations carried out by astronomers during the last four centuries (Hoyt and Schatten, 1998; Clette *et al*., 2014). Generally, programmes with long observational periods are carried out by professional observatories. The reference observatory in sunspot observations during the 20th century and the last quarter of the 19th century was the Royal Greenwich Observatory (Willis *et al*., 2013a, 2013b; 2016; Erwin *et al*., 2013). Other important historical observatories of that time related with the sunspot observations were, for example, Mount Wilson (Lefebvre *et al*., 2005), Kodaikanal (Mandal *et al*., 2017), Debrecen (Baranyi *et al*., 2016) and the Iberian observatories (Aparicio *et al*., 2014; Carrasco *et al*., 2014; Curto *et al*., 2016; Carrasco *et al*., 2018). However, there are several historical examples of single astronomers with a long observation series. For example, this is the case of Hisako Koyama with 37 observation years (Koyama, 1985; Knipp *et al*., 2017), David Hadden with 42 years (Carrasco *et al*., 2013), Sergio Cortesi with 55 years (Cortesi *et al.*, 2016; Clette *et al*., 2016), and Thomas Cragg and Herbert Luft with 63 and 65 observation years respectively (Vaquero *et al*., 2016). These long data series provided by single observers are particularly valuable in order to maintain the homogeneity of the sunspot number series, even more valuable than institutional observatories performing their observations with multiple shifts (Clette *et al*., 2014).

Eric Strach (Figure 1) was an amateur astronomer who can be included in the category of observers with a long observational series. Some biographical notes are exposed in Section 2. Section 3 contains an analysis of the Strach's records, a summary of the scientific contributions made by Strach, and the presentation of his sunspot and prominence series made during four decades. Finally, Section 4 is devoted to final comments.

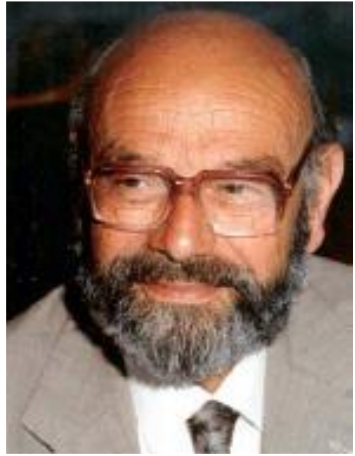

Figure 1. Eric Hugo Strach (1914-2011) [Courtesy of Journal of the British Astronomical Association, www.britastro.org].

**2. Biographical Note**

Eric Hugo Strach (1914-2011) was born in Brno (current Czech Republic) on 21 October 1914 (Ghorbal, 2011). He studied medicine at the University of Prague and graduated in 1938. When Strach finished his studies, he spent a summer in France. Due to the political situation, his parents dissuaded him to return to Czechoslovakia and he worked as a resident medical officer in Dreux, in the western part of Paris (Strach, 2012). A few years later, Strach was recruited by the Czech army and was sent to England. He could not return to Czechoslovakia until the end of the World War II. However, disillusioned by the situation of his country, he came back to England resuming his surgical career.

He became interested in astronomy and joined the Liverpool Astronomical Association and British Astronomical Association during the 1960s (Smith, 2009). Strach built an astronomical observatory in the garden of his house at Liverpool and observed the Sun when the weather allowed it. He also travelled abroad several times to observe total solar eclipses. In addition to solar observations, he was interested in the Moon and started to observe it in 1979. Strach received two awards as recognition of his solar work: the BAA's Merlin Medal and Gift in 1999 and the Walter Goodacre Medal and Gift in 2009 (Smith, 2009; Ghorbal, 2011).

**3. Analysis of the solar records made by Strach**

**3.1. General information**

The 55 observation notebooks made by Strach contain almost 10000 documents. They are composed of daily drawings and sketches of different solar activity phenomena and tables with a monthly summary indicating daily number of sunspot groups and prominences. After the death of Strach, his family donated his observation notebooks to the Historical Archive of Sunspot Observations (HASO). This archive is a special collection located in the library of the Mérida University Center (Mérida, Spain) belonging to the Extremadura University and its aim is to collect and preserve any document related with solar observations. At this moment, the Strach's documents are being digitized and will be publicly available at the HASO website (http://haso.unex.es/) when this laborious process is finished. We have assigned shelf marks to each notebook, which contain the information chronologically ordered, to facilitate to the reader the location of the information. For example, the first notebook has been referenced as "STRACH_MS1_1969-08-29_1970-07-11".

Eric Strach established his observatory (53º 23' N, 2º 52' W) at Liverpool. He used several telescopes to carry out his solar observations. According to his notebooks, for the first records, Eric observed both with a 3" ($\simeq$ 75 mm) and 4" ($\simeq$ 100 mm) refractor telescope. Afterwards, Strach acquired the 12" ($\simeq$ 300 mm) reflector owned by his friend Harold Hill, another meticulous amateur astronomer (Ghorbal, 2011). From 1979, Strach also used a 6" ($\simeq$ 150 mm) reflector telescope. Furthermore, he obtained a H-alpha filter and built his own prominence scope. Years later, Eric acquired a Daystar H-alpha narrow band filter in order to study filaments, plages, and flares. The telescope generally used by Strach during the first half of his records was the 3" telescope, projecting the Sun on a template with a 6" disc. However, Strach rarely indicated the telescope used in the second part of his observation period.

**3.2. Scientific contributions made by Strach**

Shortly after starting his observations, Strach submitted his solar records both to the Liverpool and British Astronomical Association (Ghorbal, 2011). Strach joined these associations during the 1960s. In fact, he became president of the Liverpool Astronomical Association between 1979 and 1982 (Smith, 2009). Moreover, Strach also published several scientific articles. For example, in his first contribution, Strach (1982) showed some examples of historical astronomers who were medical men as well. Later, he performed more contributions about analysis of the shadow bands recorded in the total solar eclipse of Curaçao (Netherlands Antilles) on 26 February 1998 (Strach, 1998), solar filaments and fibrils (Stach, 2002), and solar chromospheric darkenings around active areas (Strach, 2006), *inter alia*.

Strach also made scientific outreach contributions. For example, he published some booklets such as *Observing the Sun*, for beginners in astronomy, and *The Sun in Hydrogen-Alpha*, including more advanced techniques. Strach participated in observation days organized by the Liverpool Astronomical Association for the general public as the "Croxteth Park Star Party" celebrated in 2002 and attended by hundreds of people, and meetings with other astronomical associations like the Vlasim Astronomical Society (Czech Republic). He also travelled to observe 13 total solar eclipses, the first one being on the coast of Mauritania on 30 June 1973 (Ghorbal, 2011).

### 3.3. Information recorded by Strach in the sunspot drawings

Strach carried out a complete and meticulous record of solar activity including observations of sunspots, prominences, filaments, plages, and even solar flares. Figure 2 shows an example of the solar records carried out by Strach. He indicated for each drawing: date, time, seeing, instruments used (although not always, as mentioned above), number of solar rotation from Carrington rotation number, and the value corresponding to the solar parameters P, $B_0$, and $L_0$. Figure 2 (top-left panel) shows a drawing made by Strach including sunspots (black color), faculae (yellow color), and prominences (red color). Moreover, he recorded values for heliographic coordinates and Carrington longitude for individual sunspots or sunspot groups (both represented by black numbers in brackets located surrounding the solar disc), and heliographic latitudes for the prominences (red numbers in brackets). Strach also indicated the total and hemispheric number of sunspot groups and prominences. Figure 2 (top-right panel) shows another drawing made by Strach including prominences (red or black color from the solar limb outward) in addition to filaments (blue color), plages (red color inside the solar disc), and solar flares (brilliant red color inside the solar disc). Furthermore, Strach sometimes provided a detailed and expanded drawing for some regions of interest (Figure 2, bottom panel).

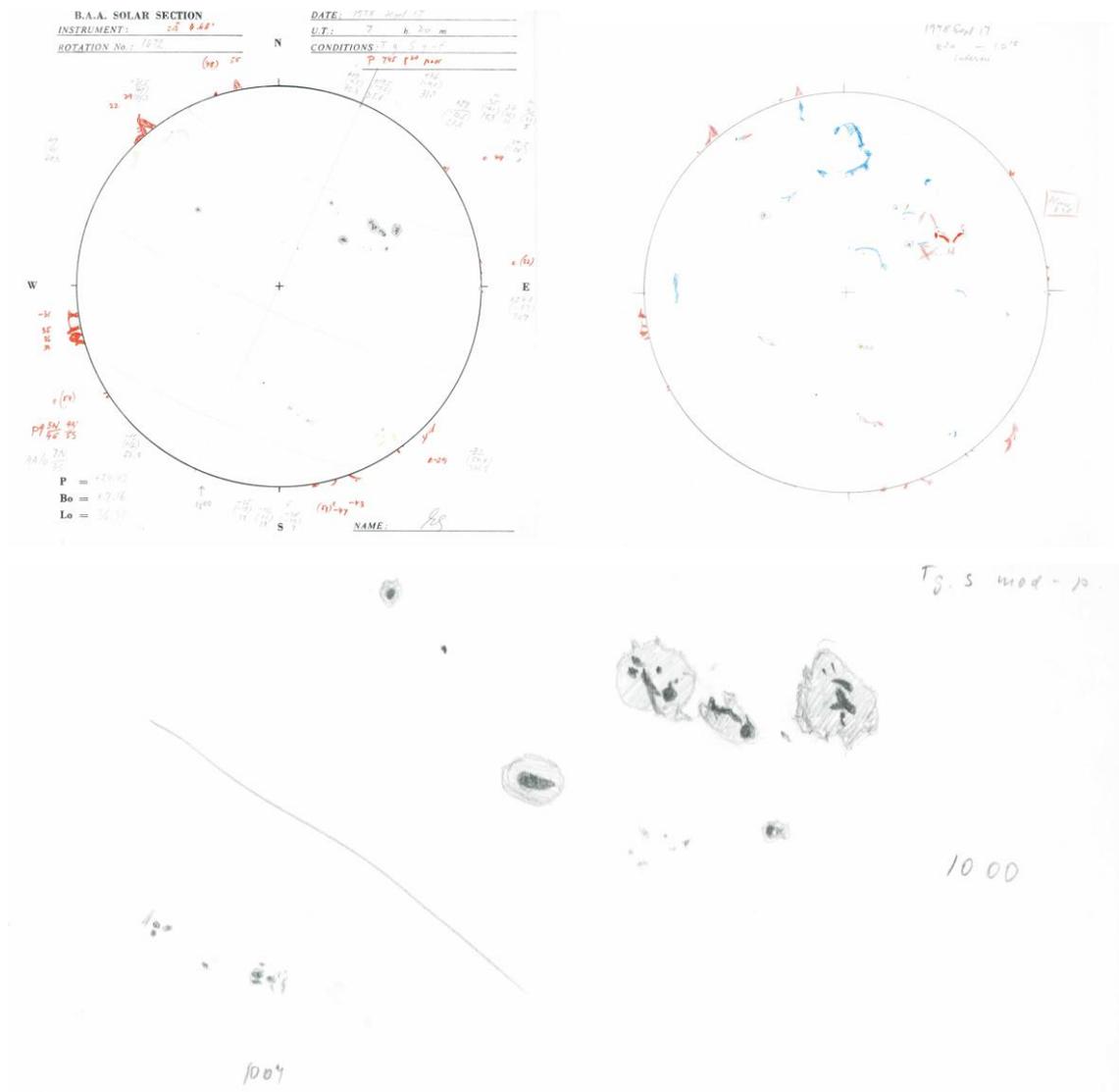

Figure 2. Original solar record made by Strach on 17 September 1978 where it is represented: i) (Top-left panel) sunspots (black), faculae (yellow), and prominences (red), together with heliographic coordinates (numbers in black color for sunspots and red color for prominences), ii) (top-right panel) prominences (red color on the solar limb), filaments (blue), plages (red color inside solar disc), and a solar flare (brilliant

red color inside solar disc), iii) and (bottom panel) a more detailed sunspot sketch made by Strach about the active region where a solar flare was observed.

**3.4. Sunspot group and prominences counting made by Strach**

At the end of each month, Strach shows in his observation notebooks a table including a summary with the daily number of sunspot groups, as it was usual in the solar section of the British Astronomical Associations, and prominences (hemispheric numbers included) recorded during the entire month. We highlight that while there is a good agreement between the sunspot counting shown in those tables and the sunspot group numbers recorded in the drawings, there is sometimes no agreement between the prominence counting shown in the tables and the prominence numbers recorded in the drawings. This fact can be due to the difficulty, for example, in establishing the number of prominences when several prominences are very close. Thus, we chose the values presented in the tables in order to analyse the Strach's records.

Sunspot group number (top panel) and prominence (bottom panel) series built from Strach's records are represented in Figure 3. Although the first solar observations made by Strach were in the late 1950s (Ghorbal, 2011), the observation notebooks only include systematic solar observations from August 1969. We must emphasize that Strach was a very constant observer with a great number of observations. Figure 3 (top panel) shows that the number of his first sunspot records was relatively low (he observed 100 days from August 1969 to December 1970). However, the annual number of sunspot observations from 1971 was always greater than 150. Strach increased his yearly average and his number of annual observations from 1975 was greater than 200, except in 1979 and 1982 when he observed 195 and 199 days, respectively. The sunspot observations made by Strach analysed in this work span from 1969 to 2008, *i.e.* four decades of sunspot records. These records started around the solar maximum of Solar Cycle 20 and finished around the minimum of the current Solar Cycle 24. The solar minima and maxima according to Strach's sunspot records lie in 1976 and 1979 (Solar Cycle 21), 1986 and 1991 (Solar Cycle 22), and 1996 and 2000 (Solar Cycle 23), respectively. The solar maxima and minima given by the sunspot number version 2 (http://www.sidc.be/silso/) are very similar to Strach's series. The only difference lies in the maximum of the Solar Cycle 22 since, according to sunspot number version 2, the solar maximum for that cycle was in 1989 (although its value, 211.1, is very similar to the value in 1991, 203.3, when it is the maximum of Solar Cycle 22 according to Strach data). The observation period of prominences made by Strach is shorter than the period corresponding to sunspot observations. Strach started the systematic prominence records in August 1974 and ended, as the case of sunspots, in 2008, *i.e.* his prominences observations span the Solar Cycles 21-23. In this case, solar maxima are in 1979, 1990, and 2000 while minima are in 1975, 1986, 1995. In general, these values of maxima and minima are not equal to those obtained from sunspot observations but they are very similar. The temporal coverage for the prominences records (49.9 %) is less than the coverage for the sunspot records (60.8 %), but we highlight that the prominences series also has a large annual observation number because only in 1974 and for the period 1978–1981 the yearly number of records is less than 150. A machine-readable version of the Strach's sunspot and prominence series is publicly available at the HASO website (http://haso.unex.es/).

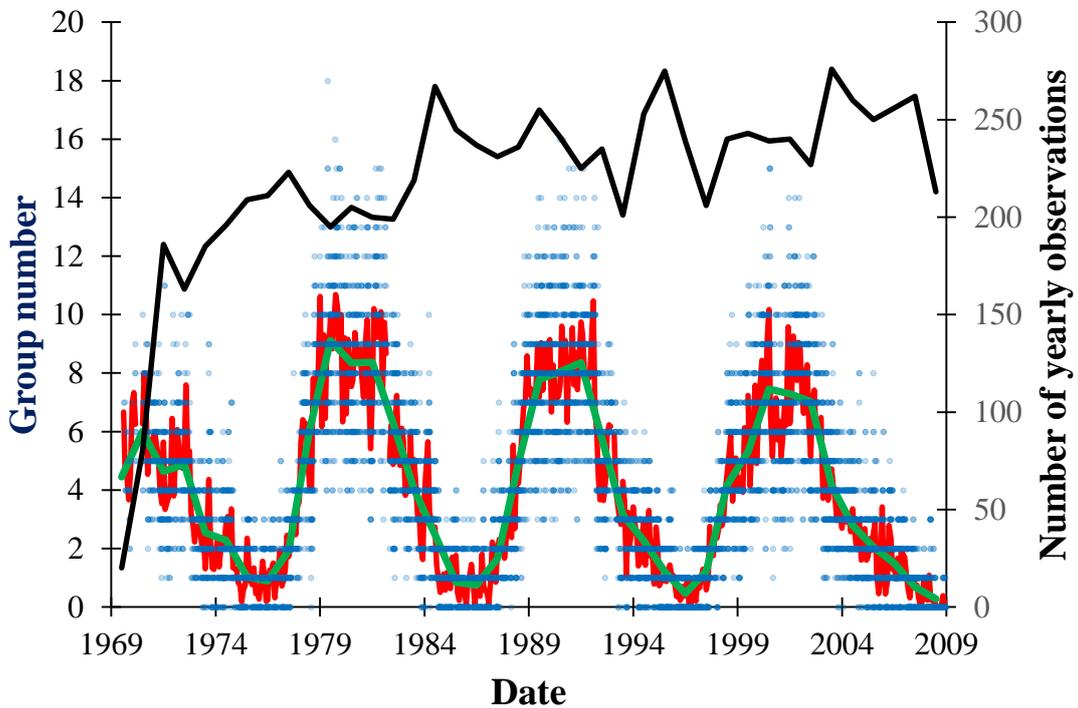

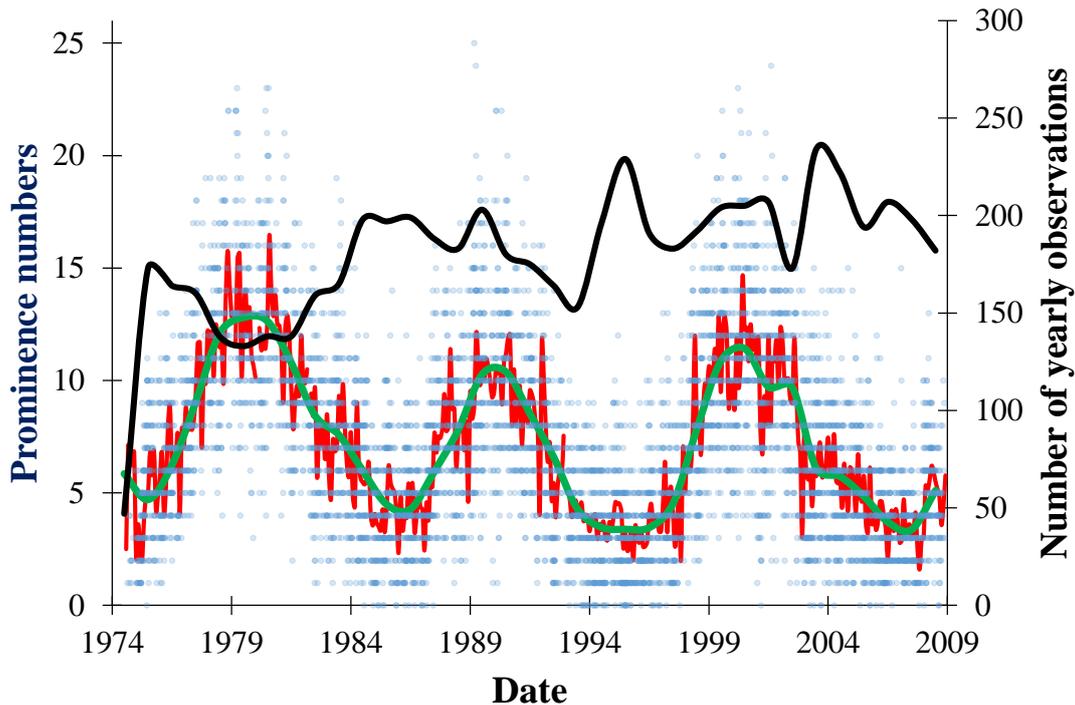

Figure 3. Sunspot group number (Top panel) and prominence (bottom panel) series built from Strach's records. The number of yearly observations made by Strach is represented by black lines. Small blue dots, and red and green lines depict the daily, monthly, and yearly values of the series, respectively.

**3.5. Filaments, plages, and solar flares observations**

Strach also carried out records of filaments, plages, and solar flares in H-alpha from 6 March 1978. He employed several filters to record these phenomena. In the first part of these records, he usually employed a 0.68 Å filter but, in the second one, he rarely indicated the filter used, although the most ones recorded in templates for this part were a 0.55 and 0.59 Å filters. Figure 4 shows an example of a solar drawing made by Strach in H-alpha (left panel) and the image taken at the Coimbra Astronomical Observatory for the same observation day (13 March 1989, http://www.astro.mat.uc.pt/novo/observatorio/site/index.html). The drawing made by Strach is very similar to the image obtained at Coimbra Observatory, especially for filaments. One of the future objectives is to analyse the filament, plage, and flare records made by Strach and to provide a machine-readable version for these phenomena.

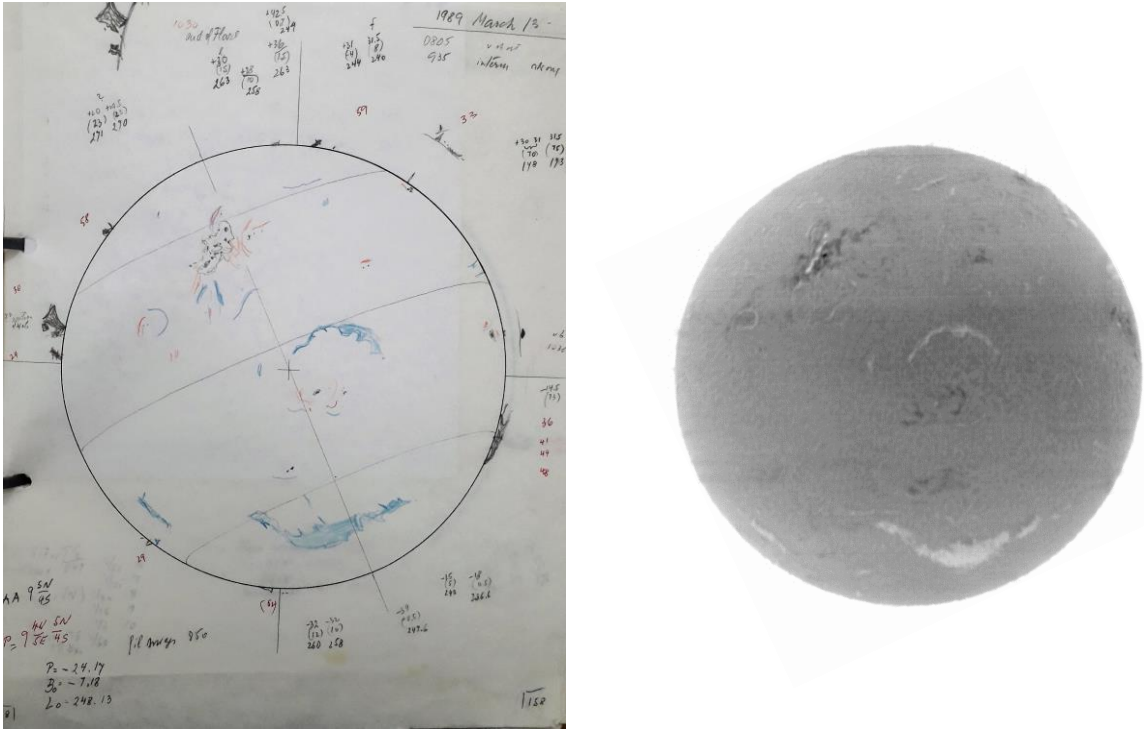

Figure 4. Original solar drawing made by Strach from a H-alpha observation (left panel) and a H-alpha image taken at Coimbra Astronomical Observatory (right panel, http://www.astro.mat.uc.pt/novo/observatorio/site/index.html), both on 13 March 1989. Note that the orientation of the Coimbra image was set in this figure to match that one of the Strach's drawing.

**5. Final comments**

In this work, we highlight the contribution of Eric Strach to solar physics. Strach devoted a large part of his life to astronomy. He joined the Liverpool Astronomical Association (where he became president from 1979 to 1982) and British Astronomical Association during the 1960s. He published several scientific articles and participated in several scientific outreach contributions such as publications for beginner astronomers and observation days organized for the general public.

Strach became an indefatigable observer carrying out meticulous and very complete solar records of sunspots, prominences, filaments, plages, and solar flares. He sometimes provided a detailed and expanded drawing for regions of interest. He started solar observations in the late of 1950s, however systematic solar records are available from August 1969 in his observation notebooks, completing four decades of solar records. Strach started his record of sunspot observations in 1969, while starting his prominence records in 1974. His yearly number of sunspot observations from 1971 was always greater than 150, achieving a great temporal coverage. In the case of prominences, only 1974 and the period 1978 – 1981 have a yearly number of observations less than 150. Thus, the temporal coverage for the sunspot records is slightly above 60 % while it is around 50 % in the prominence case. A machine-readable version of the sunspot and prominence observations is available at the HASO website. Strach also recorded filaments, plages, and solar flares from 1978. We showed that these kinds of observations carried out by Strach were very similar to observations made at professional observatories as, for example, the Coimbra Astronomical Observatory. One of our objectives for the future is that all documents included in his observation notebooks are digitized and publicly available at the HASO website.

Long and homogeneous observation series made by the same observer are very important not only for solar physics and climate studies (for example, for the ongoing reconstruction of 400-year sunspot number series) but also for space weather studies. In order to understand the behaviour of solar activity and to be able to predict it, long solar observation programmes should be continued. The space weather impact on technology or the climatic influence of solar activity were unknown in the time of Galileo and Wolf, but the efforts made by all the astronomers in the last four centuries for recording solar observations have been fundamental to understand that impact (Pevtsov and Clette, 2017). Finally, we want to recognize the work and the scientific importance of all those observers who carry out long observation series. In particular, we highlight the magnificent observation records with a four decade duration made by Eric Strach. As a future work, the digitization of all Strach's documents and analysis of filaments, plages and solar flares recorded by Strach remains to be done.

**Acknowledgements**

Data used in this work can be found at HASO website (http://haso.unex.es/). This work was partly funded by FEDER-Junta de Extremadura (Research Group Grant GR18097 and project IB16127) and from the Ministerio de Economía y Competitividad of the Spanish Government (CGL2017-87917-P). Authors have benefited from the participation in the ISSI workshops. Authors acknowledge to the British Astronomical Association the permission to use the image of the Figure 1.

**References**

Aparicio, A.J.P., Vaquero, J.M., Carrasco, V.M.S., & Gallego, M.C. (2014). Sunspot Numbers and Areas from the Madrid Astronomical Observatory (1876 – 1986), Solar Phys. 289, 4335. doi: 10.1007/s11207-014-0567-x.

Baranyi, T., Győri, L., & Ludmány, A. (2016). On-line Tools for Solar Data Compiled at the Debrecen Observatory and Their Extensions with the Greenwich Sunspot Data. Solar Physics 291, 3081. doi: 10.1007/s11207-016-0930-1.

Barnard, L., McCracken, K. G., Owens, M. J., & Lockwood, M. (2018). What can the annual 10Be solar activity reconstructions tell us about historic space weather? Journal of Space Weather and Space Climate 8, A23. doi: 10.1051/swsc/2018014.

Carrasco, V.M.S., Vaquero, J.M., Gallego, M.C., & Trigo, R.M. (2013). Forty two years counting spots: solar observations by DE Hadden during 1890–1931 revisited, New Astron. 25, 95. doi: 10.1016/j.newast.2013.05.002.

Carrasco, V.M.S., Vaquero, J.M., Aparicio, A.J.P., & Gallego, M.C. (2014). Sunspot Catalogue of the Valencia Observatory (1920 – 1928), Solar Phys. 289, 4351. doi: 10.1007/s11207-014-0578-7.

Carrasco, V.M.S., Vaquero, J.M., Gallego, M.C, Lourenço, A., Barata, T., & Fernandes, J.M. (2018). Sunspot Catalogue of the Observatory of the University of Coimbra (1929–1941), Solar Phys. 293, 153. doi: 10.1007/s11207-018-1373-7.

Clette, F., Svalgaard, L., Vaquero, J.M., & Cliver, E.W. (2014). Revisiting the Sunspot Number. A 400-year perspective on the solar cycle, Space Sci. Rev. 186, 35. doi: 10.1007/s11214-014-0074-2.

Clette, F., Lefèvre, L., Cagnotti, M., Cortesi, S., & Bulling, A. (2016). The Revised Brussels-Locarno Sunspot Number (1981 - 2015). Solar Physics, 291, 2733. doi: 10.1007/s11207-016-0875-4.

Cliver, E.W., & Keer, N.C. (2012). Richard Christopher Carrington: Briefly Among the Great Scientists of His Time, Solar Physics, 280, 1. doi: 10.1007/s11207-012-0034-5.

Cortesi, S., Cagnotti, M., Bianda, M., Ramelli, R., & Manna, A. (2016). Sunspot Observations and Counting at Specola Solare Ticinese in Locarno Since 1957, Solar Physics 291, 3075. doi: 10.1007/s11207-016-0872-7.

Curto, J.J., Solé, J.G., Genescà, M., Blanca, M.J., & Vaquero, J.M. (2016). Historical Heliophysical Series of the Ebro Observatory, Solar Phys. 291, 2587. doi: 10.1007/s11207-016-0896-z

Erwin, E.H., Coffey, H.E., Denig, W.F., Willis, D.M., Henwood, R., & Wild, M.N. (2013). The Greenwich Photo-heliographic Results (1874 – 1976): Initial Corrections to the Printed Publications, Solar Phys. 288, 157. doi: 10.1007/s11207-013-0310-z.

Ghorbal, M. (2011). Obituary: Eric Hugo Strach, 1914-2011, *J. Br. Astron. Assoc.* 121, 5.

Hayakawa, H., Iwahashi, K., Ebihara, Y., Tamazawa, H., Shibata, K., Knipp, D.J., Kawamura, A.D., Hattori, K., Mase, K., Nakanishi, I., Isobe, H. (2017). Long-lasting Extreme Magnetic Storm Activities in 1770 Found in Historical Documents, The Astrophysical Journal Letters 850, L31. doi: 10.3847/2041-8213/aa9661.

Hayakawa, H., Ebihara, Y., Hand, D.P., *et al*. (2018). Low-latitude Aurorae during the Extreme Space Weather Events in 1859. The Astrophysical Journal, 869, 57. doi: 10.3847/1538-4357/aae47c.

Hayakawa, H., Ebihara, Y., Cliver, E.W., *et al*. (2019). The extreme space weather event in September 1909. Monthly Notices of the Royal Astronomical Society, 484, 4083. doi: 10.1093/mnras/sty3196.

Hoyt, D.V., & Schatten, K.H. (1998). Group Sunspot Numbers: A New Solar Activity Reconstruction, Solar Phys. 179, 189. doi: 10.1023/A:1005007527816.

Ishii, T.T., Kurokawa, H., & Takeuchi, T.T. (1998). Emergence of a Twisted Magnetic Flux Bundle as a Source of Strong Flare Activity. The Astrophysical Journal, 499, 2, 898. doi: 10.1086/305669.

Knipp, D., Liu, H., & Hayakawa, H. (2017). Ms. Hisako Koyama: From amateur astronomer to long-term solar observer. Space Weather, 15, 1215. doi: 10.1002/2017SW001704.

Koyama, H. (1985). Observations of sunspots, 1947 – 1984, Kawadeshoboshinsha, Tokyo.

Lefebvre, S., Ulrich, R.K., & Webster, L.S. (2005). The solar photograph archive of the Mount Wilson Observatory. A resource for a century of digital data, Mem. S.A.It. 76, 862.

Lefèvre, L., Vennerstrøm, S., Dumbović, M., Vršnak, B., Sudar, D., Arlt, R., Clette, F., Crosby, N. (2016) Solar Physics 291, 1483. doi: 10.1007/s11207-016-0892-3.

Love, J.J. (2018). The Electric Storm of November 1882, Space Weather 16, 37. doi: 10.1002/2017SW001795.

Mandal, S., Hegde, M., Samanta, T., Hazra, G., Banerjee, D., & Ravindra, B. (2017). Kodaikanal digitized white-light data archive (1921-2011): Analysis of various solar cycle features, Astron. Astrophys. 601, A106. doi: 10.1051/0004-6361/201628651.

Muscheler, R., Adolphi, F., Herbst, K., & Nilsson, A. (2016). The Revised Sunspot Record in Comparison to Cosmogenic Radionuclide-Based Solar Activity Reconstructions. Solar Physics, 291, 3025. doi: 10.1007/s11207-016-0969-z.

Pevtsov, A.A., & Clette, F. (2017). To understand future solar activity, one has to know the past, Eos, 98. doi: 10.1029/2017EO083277.

Riley, P., Baker, D., Liu, Y.D., Verronen, P., Singer, H., & Güdel, M. (2018). Extreme Space Weather Events: From Cradle to Grave, Space Sci. Rev. 214, 21. doi: 10.1007/s11214-017-0456-3.

Saiz, E., Cid, C., Guerrero, A. (2018). Environmental Conditions During the Reported Charging Anomalies of the Two Geosynchronous Satellites: Telstar 401 and Galaxy 15, Space Weather, doi: 10.1029/2018SW001974.

Smith, L. (2009). A tribute to Eric Strach - one of our great amateur observers, J. Br. Astron. Assoc. 119, 292.

Strach, A. (2012). Obituary: Eric Hugo Strach, BMJ 344, 1419. doi: 10.1136/bmj.e1419.

Strach, E. (1982). Astronomy and Medicine, J. Br. Astron. Assoc. 92, 164.

Strach, E. (1998). Shadow bands recorded at February 26 eclipse, J. Br. Astron. Assoc. 108, 127.

Strach, E. (2002). Filaments and fibrils on the Sun, J. Br. Astron. Assoc. 112, 229.

Strach, E. (2006). Solar chromospheric darkenings around active areas, J. Br. Astron. Assoc. 116, 19.

Toriumi, S., Schrijver, C.J., Harra, L. K., Hudson, H., & Nagashima, K. (2017). Magnetic Properties of Solar Active Regions That Govern Large Solar Flares and Eruptions, Astrophys. J. 834, id. 56. doi: 10.3847/1538-4357/834/1/56.

Usoskin, I.G. (2017). A history of solar activity over millennia, Living Rev. Solar Phys. 14, 3. doi: 10.1007/s41116-017-0006-9.

Usoskin, I.G., Mursula, K., Solanki, S., Schüssler, M., & Kovaltsov G.A. (2002). A physical reconstruction of cosmic ray intensity since 1610. Journal of Geophysical Research (Space Physics), 107, A11. doi: 10.1029/2002JA009343.

Vaquero, J.M., Svalgaard, L., Carrasco, V.M.S., Clette, F., Lefèvre, L., Gallego, M.C., Arlt, R., Aparicio, A.J.P., Richard, J.-G., & Howe, R. (2016). A Revised Collection of Sunspot Group Numbers, Solar Phys. 291, 3061. doi: 10.1007/s11207-016-0982-2.

Willis, D. M., Wild, M. N., & Warburton, J. S. (2016). The Greenwich Photo-heliographic Results (1874 - 1885): Observing Telescopes, Photographic Processes, and Solar Images. Solar Physics 291, 2519. doi: 10.1007/s11207-016-0856-7.

Willis, D.M., Coffey, H.E., Henwood, R., Erwin, E.H., Hoyt, D.V., Wild, M.N., & Denig, W.F. (2013a). The Greenwich Photo-heliographic Results (1874 – 1976): Summary of the Observations, Applications, Datasets, Definitions and Errors, Solar Phys. 288, 117. doi: 10.1007/s11207-013-0311-y.

Willis, D.M., Henwood, M.N., Wild, M.N., Coffey, H.E., Denig, W.F., Erwin, E.H., & Hoyt, D.V. (2013b). The Greenwich Photo-heliographic Results (1874 – 1976): Procedures for Checking and Correcting the Sunspot Digital Datasets, Solar Phys. 288, 141. doi: 10.1007/s11207-013-0312-x.